\pdfoutput=1
\documentclass[conference]{IEEEtran}
\usepackage[utf8]{inputenc}
\usepackage{enumerate}
\usepackage{hyperref}
\usepackage[tight]{subfigure}
\usepackage{graphicx}
\usepackage{amsmath}

\title{Measurement of single event upsets in the ALICE-TPC front-end electronics}

\author{M.~Mager, L.~Musa, A.~Rehman, A.~Szczepankiewicz for the ALICE-TPC collaboration\\
  CERN
}

\begin{document}
\bstctlcite{IEEEexample:BSTcontrol}
\maketitle
\begin{abstract}
  \boldmath
  The Time Projection Chamber of the ALICE experiment at the CERN Large Hadron Collider features highly integrated on-detector read-out electronics. It is following the general trend of high energy physics experiments by placing the front-end electronics as close to the detector as possible---only some 10\;cm away from its active volume. Being located close to the beams and the interaction region, the electronics is subject to a moderate radiation load, which allowed us to use commercial off-the-shelf components. However, they needed to be selected and qualified carefully for radiation hardness and means had to be taken to protect their functionality against soft errors, i.\,e.\ single event upsets.

Here we report on the first measurements of LHC induced radiation effects on ALICE front-end electronics and on how they attest to expectations.
\end{abstract}


\section{Introduction}

\begin{figure}[b]
  \centering
  \includegraphics[angle=90,width=7cm]{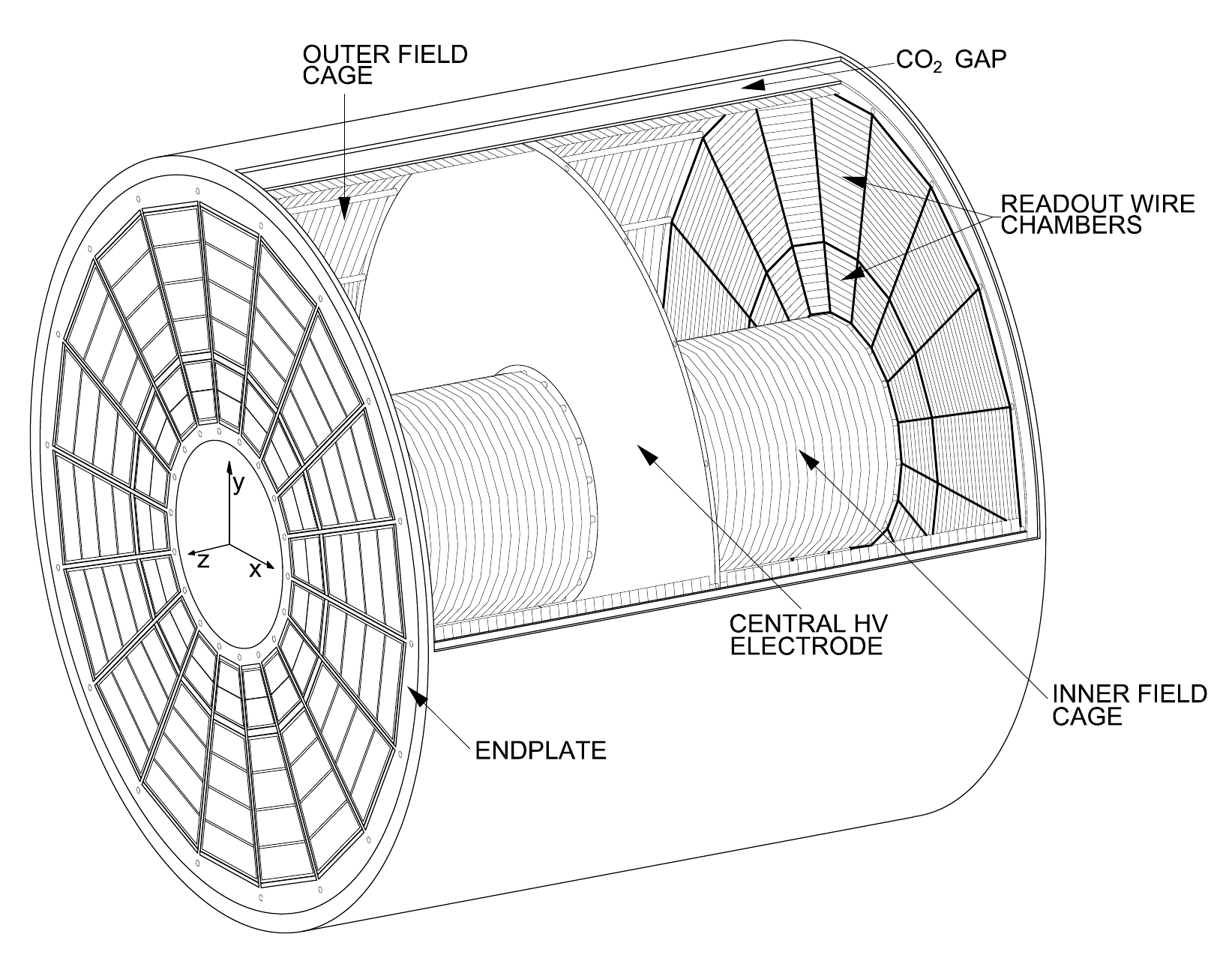}
  \caption{Layout of the TPC (picture from \cite{TPC-NIM}).}
  \label{fig:tpc}
\end{figure}

\begin{table}[b]
  \caption{Distribution of the front-end electronics and bits masked in the analysis.}
  \label{tab:partitions}
  \centering
  \begin{tabular}{cccrl}
    \hline
    Partition & Radial position            & FECs/sector & \multicolumn{2}{c}{Bits/TPC (masked)} \\
    \hline
    0 & 100\;cm & 18 & $8.49\cdot10^{8}$ & ($2.88\cdot10^{7}$) \\
1 & 120\;cm & 25 & $1.18\cdot10^{9}$ & ($4.20\cdot10^{7}$) \\
2 & 149\;cm & 18 & $8.49\cdot10^{8}$ & ($1.31\cdot10^{6}$) \\
3 & 175\;cm & 20 & $9.44\cdot10^{8}$ & ($0$) \\
4 & 203\;cm & 20 & $9.44\cdot10^{8}$ & ($0$) \\
5 & 228\;cm & 20 & $9.44\cdot10^{8}$ & ($1.02\cdot10^{4}$) \\

    \hline
  \end{tabular}
\end{table}
Radiation induced effects on the electronics of high energy physics experiments are gaining importance due to the ever increasing density of electronics and their placement into zones of high particle flux. When energetic particles cross the electronics they may release a significant amount of charge along their paths leading to diverse erratic behaviours, termed single event effects (SEEs). Amongst them are single event upsets (SEUs), which refer to flip-flops/memory cells changing their state ($0\to1$ or $1\to0$). These effects are of major importance, even in areas of moderate radiation load, because they may accumulate and lead to system failures if no mitigation is applied. The estimation of SEU rates and the protection of critical bits is therefore an important task in the design process of detector read-out electronics.

The ALICE Time Projection Chamber (TPC) is the main tracking detector of the ALICE experiment at the CERN Large Hadron Collider (LHC) \cite{TPC-NIM,TPC-TDR}. It comprises a hollow cylinder, placed concentrically around the beam-line, with an active volume constrained by $84.8<r<246.6\;\text{cm}$ (radial) and $\lvert z\rvert<249.7\;\text{mm}$ (along the beam line, see Fig.~\ref{fig:tpc}). The detector is equipped with multi-wire proportional chambers that are mounted on the end-plates and uses a high granularity pad read-out with 557,568 pads.

The large number of active channels required to mount the electronics for digital pre-processing and temporary data storage onto the detector; it is placed only some 10\;cm away from the end-plates. The signal amplification, digitisation and processing is performed by two custom ASICs: the pre-amplifier and shaping amplifier (PASA, 0.35\;$\mu$m) and the ADC and signal processor (ALTRO, 0.25\;$\mu$m). A total of 34,848 ALTROs and 34,848 PASAs are utilised to read out the detector.

While designing electronics to be operated that close to the interaction region, radiation tolerance has been anticipated to be of importance since the very first design stages. Detailed simulations showed that, at its position, the radiation load on the electronics is on the one hand too high to use random off-the-shelve electronics, but on the other hand too low to justify the cost of radiation hardened components. Instead, lots of effort was put on a) qualifying commercial and custom made components for their radiation hardness in terms of total dose and b) implement mitigation for single event effects \cite{Roed-PhD,ALICE-INT-2002-028}.

Here we report on the first direct observation of SEUs in the TPC front-end electronics, i.\,e.\ in the ALTRO memories, which are attributed to particles emerging from collisions. At this early stage of operation of accelerator and experiment, priority is given to understanding the detector signals in detail. This in particular implies that not all sophisticated features of the electronics for on-detector signal processing are switched on, which in turn allowed us to utilise them as a radiation monitor. It should be noted that the measurement thus does not interfere with the normal operation and data-taking of the detector at any stage.
\begin{table*}[t]
  \caption{Parameters of selected subsystems and their mitigation strategy.}
  \label{tab:scaling}
  \centering
  \begin{tabular}{ccccc}
    \hline
    Component & Part number            & Bits/TPC       & SEU cross-section                                  & Mitigation \\
    \hline
    ALTRO     & ---                   & $29\cdot10^9$  & $5.5\cdot10^{-14}\;\text{cm}^2\text{bit}^{-1}$ & Hamming-encoded state machines\\
    BC        & Altera AP1K30TC144-3  & $0.92\cdot10^9$ & unknown & none \\
    RCU       & Xilinx XC2VP7-6-FF672 & $2.0\cdot10^9$  & $3.7\cdot10^{-14}\;\text{cm}^2\text{bit}^{-1}$ & active partial reconfiguration \\
    \hline
  \end{tabular}
\end{table*}

The electronics is mounted on 4,356 front-end cards (FECs), each one housing 8 ALTROs and 8 PASAs. They are distributed over 2 sides (``A'' and ``C''), 18 sectors in azimuthal, and 6 partition in radial direction (Fig.~\ref{fig:tpc}). The number of FECs per partition is determined by the anticipated track density and the trapezoidal shape of the sectors (Tab.~\ref{tab:partitions}). This vast amount of ``unused'' bits and spatial distribution of the electronics makes the TPC a unique device for a quantitative differential measurement of hadron flux.

Apart from the read-out chips themselves, logic is added to monitor the physical parameters (temperatures, voltages, and currents) of each FEC (``board controller'', BC), as well as, to steer the read-out process of each partition by a read-out control unit (RCU). These devices are based on SRAM FPGA and thus sensitive to SEUs as well. Although here we do not measure the SEUs in those, we are able to infer their number by scaling the measured cross-sections and number of sensitive bits.

\section{Anticipated radiation effects}

\subsection{Radiation environment}\label{sec:anticipated_radiation}
Extensive simulations of both the instantaneous flux and the integrated fluency for different particle species were carried out \cite{ALICE-INT-2002-028}. Here we follow a simplified model, by assuming a flat pseudo-rapidity distribution $\text{d}{N_{ch}}/\text{d}\eta$ of primary charged particles $N_{ch}$ over the relevant values of pseudo-rapidity $\eta$. For a given radial position $r$ on the end-plate we obtain their flux (number of charged particles $N_{ch}$ per area $S$) as:
\begin{equation}\label{eq:flux}
  \frac{\text{d}{N_{ch}}}{\text{d}^2S}(r)=\frac{1}{2\pi}\cdot\frac{1}{r^2\sqrt{1+(r/Z_0)^2}}\cdot\frac{\text{d}N_{ch}}{\text{d}\eta}(\eta=0)\;,
\end{equation}
where $Z_0=\pm263\;\text{cm}$ is the $z$-position of the respective ALTRO chips. Furthermore we assume that the number of particles that cause SEUs is proportional via a factor $A$ to this number. We hereby neglect any effect due to beam--gas interactions or precise treatment of secondary particle production. Also, the containment of low energetic particles due to the magnetic field (solenoidal, 0.5\;T in beam direction) is disregarded.

Finally, the number of SEUs is proportional to the particle track length per volume rather than to the number of particles per area, which introduces another factor of $\sqrt{1+(r/Z_0)^2}$. This yields our final model:
\begin{align}\label{eq:nseu}
  N_{SEU}(r)&=A\cdot\frac{\text{d}{N_{ch}}}{\text{d}^2S}(r)\cdot\sigma_{SEU}\cdot\sqrt{1+(r/Z_0)^2}\nonumber\\
  &=\frac{A}{2\pi}\cdot\frac{1}{r^2}\cdot\sigma_{SEU}\cdot\frac{\text{d}N_{ch}}{\text{d}\eta}(\eta=0)\;.
\end{align}

The tests had been conducted with 7\;TeV minimum bias proton-proton collisions for which one obtains $\text{d}N_{ch}/\text{d}\eta(\eta=0)\approx 6$ \cite{ALICE-dNdy7TeV}. Running with heavy ions one assumes a factor of 100 more primary particles per collision \cite{Armesto-2008}. In its final configuration the LHC will deliver proton-proton and heavy-ion collisions with approximate rates of up to 1\;MHz and 10\;kHz, respectively. Both scenarios lead to mean time between failures (the crucial parameter for mitigation strategies) of about a factor 10 smaller as compared to the current running conditions.

Also very simplistic, the model correctly describes the qualitative behaviour and the order of magnitude of the simulations. For the sake of simplicity and because the overall uncertainties of our measurement fall within that range, we follow this model.

\subsection{SEU cross-sections}
All components of the front-end read-out were characterised in test beams \cite{Roed-2005,ALTRO-IEEE} and qualified for radiation hardness. The measured SEU cross-section for the ALTRO memories was obtained to be $5.5\cdot10^{-14}\;\text{cm}^2\text{bit}^{-1}$. A table summarising the number of bits and the respective cross-sections for the other selected devices in the read-out chain is given in Tab.~\ref{tab:scaling}. Moreover the components were qualified in terms of total dose. Here only components withstanding a total dose of 20 times larger than the anticipated one were accepted in order to take into account the simulation uncertainties of total fluency.

\subsection{Implemented mitigation techniques}
Since physical shielding of the electronics against radiation is not possible for the TPC due to the layout of the detector, any radiation-related errors in the read-out electronics system need to be handled by mitigation techniques at architecture and circuit level. Depending on the anticipated likelihood of errors and their possible system-level impact, different  mitigation techniques were chosen for the different components. This is summarised in Tab.~\ref{tab:scaling} and discussed hereunder.

At the predicted error rates (below 1 bit per second), data corruption is of no big concern. The used data transfer protocol ensures that errors do not spread, such that at most one single channel in one single event gets affected by an SEU. The ALTRO chip, however, implements protection in the memory management and interface state machines. They are the most important state machines dealing with the on-chip data buffers and interface protocol respectively and any failure can lead to scrambling event data streams or even electrical clashes on the bus, irreversibly damaging the electronics. These finite state machines are protected against SEU based on Hamming-encoding of the state vectors. They are thus protected against effects of single SEUs (one bit-flip) and may report the case of a double bit-flip. In the latter case the correct state vector may not be recovered and the state machine goes back to idle state. Any Hamming error is tracked and reported in a status register.

Another weak point are the configuration registers of the ALTRO. Their corruption can cause the data integrity problems of single chips. Systematic reconfiguration during the data taking process of these registers is foreseen to keep such errors very localised in time.

The FEC's board controller is implemented in a SRAM based FPGA. It is programmed from an on-board flash device at power up. In case of any malfunction due to radiation effects, it can be reprogrammed from the on board flash device to restore the configuration with a software command.

The event read-out of each partition is steered by a read-out control unit (RCU), which is implemented in another SRAM based FPGA. Active partial re\-con\-fi\-gu\-ra\-tion is employed to protect its configuration memory. It is realised by using of a flash based support FPGA that communicates both with the selectMAP interface of the Xilinx device and to an on-board flash memory device. The latter stores needed configuration files for the Xilinx.

\section{SEU measurements}
\subsection{Procedure}
The measurement of SEUs, on which we report here, employs currently unused memory cells of the ALTRO pedestal memories by monitoring bit flips, which is a widely adopted method in dedicated radiation monitoring systems (e.\,g.~\cite{Makowski-2007}).
The pedestal memories consist of 1024 10-bit SRAM cells for every ALTRO channel. There are 16 channels in each of the 34,848 ALTRO chips accounting for a total number of $5.7\cdot10^9\;\text{bit}$ available to monitor SEUs.

\begin{table}[t]
  \caption{Data sets with at least one SEU. }
  \label{tab:datasets}
  \centering
  \begin{tabular}{ccccc}
    \hline
    Start (UTC) & Duration & Collisions & SEUs\\
    \hline
    08/10/2010 18:04 & 42.4 h & $1.09\cdot10^{6}$ & 2 \\
10/10/2010 15:32 & 9.0 h & $1.43\cdot10^{7}$ & 4 \\
11/10/2010 00:30 & 7.7 h & $3.40\cdot10^{6}$ & 4 \\
11/10/2010 08:10 & 0.7 h & $0$ & 1 \\
11/10/2010 13:07 & 18.5 h & $9.15\cdot10^{8}$ & 186 \\
12/10/2010 07:35 & 10.2 h & $0$ & 1 \\
12/10/2010 22:59 & 25.6 h & $1.36\cdot10^{2}$ & 7 \\
14/10/2010 02:17 & 7.6 h & $1.63\cdot10^{9}$ & 206 \\
15/10/2010 07:00 & 12.4 h & $9.61\cdot10^{7}$ & 8 \\
15/10/2010 19:52 & 8.4 h & $3.15\cdot10^{8}$ & 19 \\
16/10/2010 04:19 & 35.1 h & $2.80\cdot10^{9}$ & 324 \\

    \hline 
  \end{tabular}
\end{table}


The measurement was carried out as follows:
\begin{enumerate}[(a)]
\item The memories were initialised with specific bit patterns (either ``0101010101'' or ``1111111111'').
\item Right after being initialised, the contents of the memories were verified against the used pattern. This allowed to detect faulty parts of the system.
\item The detector was operated for some time.
\item The memories were read back and all locations were checked against their reference data obtained in step (b).
\end{enumerate}
A list of the conducted measurements is given in Tab.~\ref{tab:datasets}. 
Their timing is constrained by the operation of accelerator and detector. A write and read cycle takes about 10 minutes and within that time no data taking may happen. Because beam time is precious, it was performed whenever there was a gap right before injection and right after dumping the beam.

Because the anticipated number of errors is very low, a careful selection of ``good'' memories was essential for obtaining a high sensitivity and remove false-positives in the SEU detection. We had to mask channels for different reasons:
\begin{itemize}
\item Single FECs were not operational due to technical problems and needed to be kept off.
\item The memories of a few channels show erratic behaviour and reported random data.
\item Two partitions had to be excluded due to a too large number of bad channels and communication problems.
\end{itemize}
A summary of available and masked bits per partition is given in Tab.~\ref{tab:partitions}.

The number of SEUs is correlated to the total number of collisions that happened during a measurement interval. This number is obtained by integrating the instantaneous collision rate as reported by the V0 detector of ALICE. It is run continuously and its values are reported to the LHC-wide database.

\subsection{Analysis}
To analyse the results in a quantitative way, all data sets were normalised to the respective number of concerned bits. This takes into account the number of masked channels, which were the same throughout the analysis of all datasets.

\begin{figure}[t]
  \centering
  \includegraphics[scale=0.75]{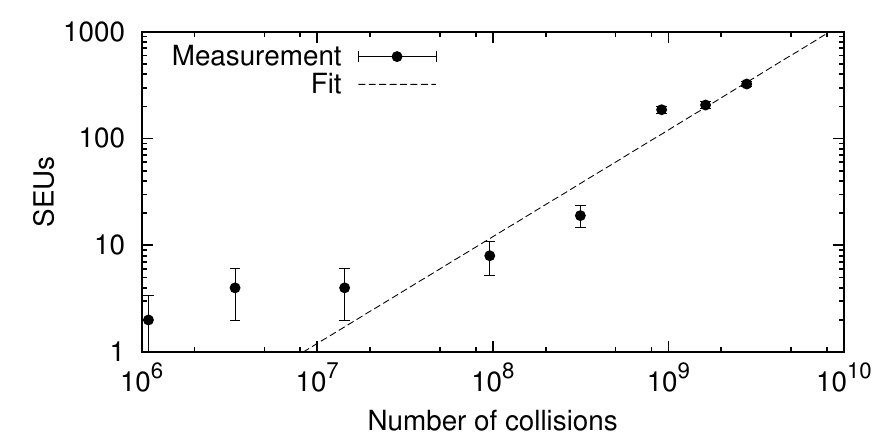}
  \caption{Correlation of integrated number of collisions and number of SEUs within a measurement period and fitted proportionality. The error bars correspond to $1\sigma$ counting errors.}
  \label{fig:vstime}
\end{figure}

The dependence of the number of SEUs on the number of collisions is shown in Fig.~\ref{fig:vstime}. It indicates proportionality and thus gives first the indication that SEUs are the origin of the observed bit errors. The proportionality constant of $(1.2\pm0.1)\cdot 10^{-7}$ per collision can directly be translated into an error frequency when multiplied by the collision rate.

\begin{figure*}[t]
  \centering
  \subfigure[Accumulated single event upsets (black boxes indicate partitions not analysed). Disabled FECs or bits ($<10\%$) are not taken into account.]{\includegraphics[scale=0.75]{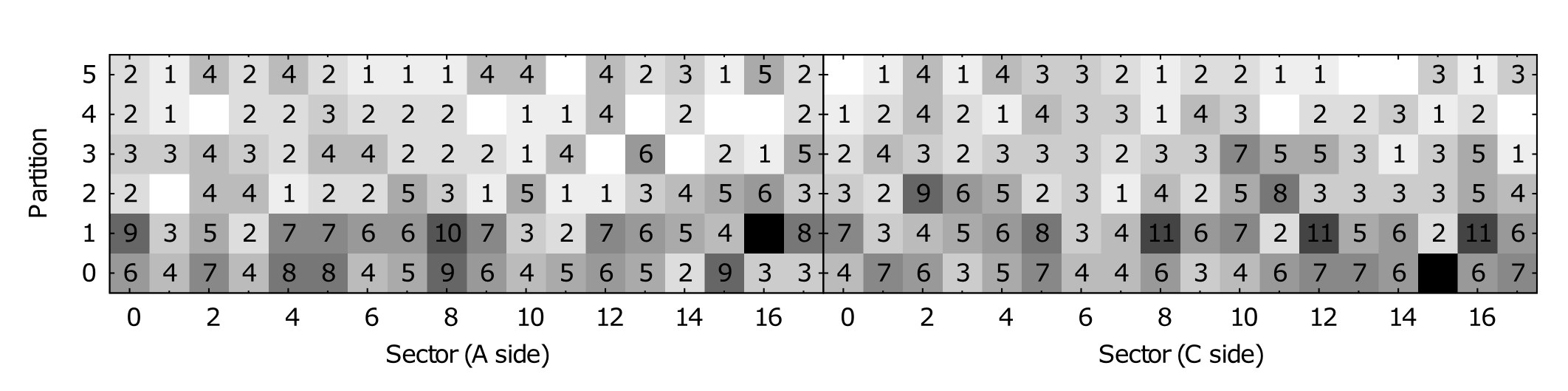} }
  \\
  \subfigure[Normalised differential distributions.]{\includegraphics[scale=0.75]{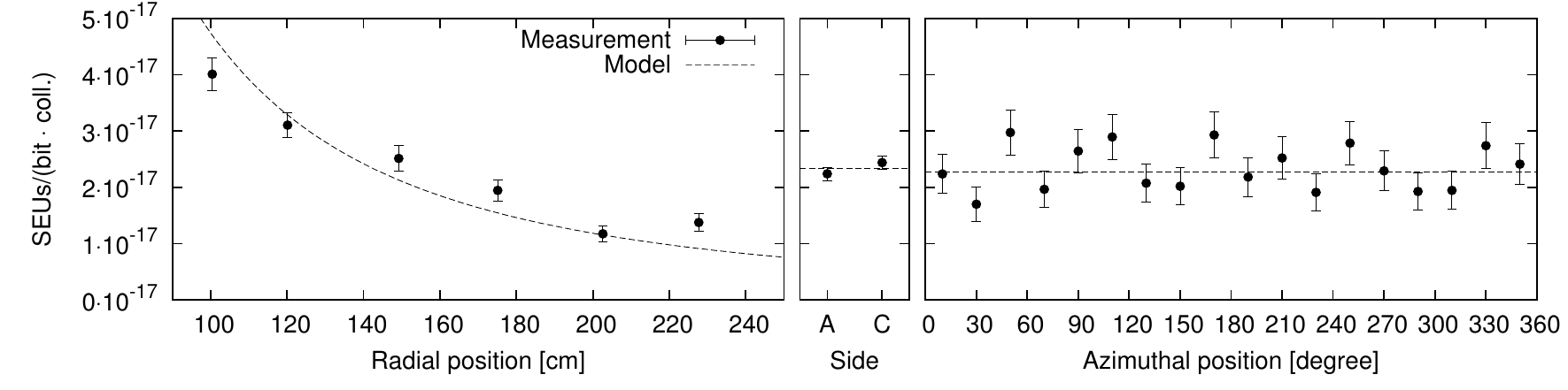}}
  \caption{Accumulated single event upsets (integral over all measurements). The error bars correspond to $1\sigma$ counting errors.}
  \label{fig:measurement}
\end{figure*}

Next, we analysed the spatial dependence of the effect, which is predicted by Eq.~\ref{eq:nseu}. The result is depicted in Fig.~\ref{fig:measurement}, which shows the integrated number of SEUs over all data sets. The absolute number of errors within each read-out partition, as well as normalised SEU probabilities per bit and collision, differential in radial direction and azimuthal angle, are provided.

Albeit limited statistics, the data indicates a radial dependence (see Fig.~\ref{fig:measurement}(b)) as it is expected from a beam induced process (Eq.~\ref{eq:nseu}). Also, Fig.~\ref{fig:measurement}(b) shows no correlation between the number SEUs and sector position, i.\,e.\ no azimuthal dependence, as expected. Even the indicated slight asymmetry between A- and C-side is expected from simulations and are due to a muon absorber placed in the C-side.

Taking the cross-section from Tab.~\ref{tab:scaling} we obtain the proportional parameter $A$ of Eq.~\ref{eq:nseu} to be $4.5\pm0.3$, which, taken into consideration the discussion in Sec.~\ref{sec:anticipated_radiation}, is very reasonable.

To finally rule out the possibility that the observed effects were not due to collision induced SEUs, but e.\,g.\ due to some electrical or thermal problem, the same measurement procedure was repeated at a technical stop (i.\,e.\ the LHC was off and thus did not deliver any beam). Thereby the operating conditions were chosen as close as possible to the situation with bit errors: the TPC was operated and took (noise) data at similar rates. In these conditions no errors were observed.

\section{Conclusions}
We have presented the first quantitative measurement of single event upsets with ALICE. It was shown to comply with theoretical expectations quantitatively within an order of magnitude and qualitatively by resembling the radial dependence. 

The measurement does not only justify the effort put on designing the built-in protection of the system but also underlines its importance in view of the anticipated increase of particle flux in heavy-ion and high rate proton-proton collisions at LHC. Moreover, it clearly shows that already at this early stage of LHC operation SEUs became visible and should not be neglected.

\section*{Acknowledgements}
The authors acknowledge the support and stimulating discussions with Andreas Morsch and Ketil R{\o}ed as well as the help of the ALICE shift crew in carrying out the measurements.

\bibliographystyle{IEEEtran}
\bibliography{seus.bib}

\end{document}